# Feasibility of backward diffraction radiation for non-destructive diagnostics of relativistic charged particle beams.

A.P.Potylitsyn, N.A.Potylitsyna

Tomsk Polytechnic University, pr. Lenina 30, 634004 Tomsk, Russia

**Abstract**

The characteristics of backward diffraction radiation (BDR), i.e. radiation of the charged particle passing through a slit in the tilted screen, has been considered. The technique for non-destructive beam diagnostics based on the measurements of BDR yield for different tilted angles (theta-scan) is suggested.

PACS numbers: 41.60.m, 41.75.i

Keywords: diffraction radiation, beam diagnostics.

Diffraction radiation (DR) is generated when a charged particle rectilinearly moves in the vicinity to a conducting screen (target) in vacuum. As in the case of transition radiation (TR), appearing when a particle crosses a tilted conducting plane, DR characteristics are determined by both the Lorentz – factor of the particle γ and its charge. Also, they do not depend on the mass of the particle. In the paper [1] the exact solution of the Maxwell equations was obtained, describing radiation of charged particle passing above a tilted semi-infinite ideally conducting screen. One of the authors of the present paper showd [2] that for ultrarelativistic particles ($\gamma \gg 1$) the DR characteristics for the geometry considered are quite close to characteristics of TR.

Optical transition radiation (OTR) is known to be widely used for charged particle beam diagnostics [3,4]. In the paper [5] M. Castellano considered a possibility of a transverse beam size determination using optical diffraction radiation characteristics when a beam moves through a slit in a perpendicular screen. In the quoted paper it was suggested to measure the DR angular distribution in the vicinity to the initial electron beam direction. In analogy with transition radiation one can talk about "forward diffraction radiation, FDR". For this case, one should notice that the beam diagnostics is carried out using "backward transition radiation, BTR", i.e. the radiation from a tilted foil in the vicinity of the specular direction is investigated because in that case the detector can be placed far enough from the beam for soft background conditions. In analogy with BTR one can expect that in case when relativistic particle moves through a slit in a tilted screen, the radiation is generated along the specular direction which can be called as "backward diffraction radiation, BDR" [6]. In the present paper the BDR characteristics of an ultrarelativistic particle when it moves through a slit in an ideally conducting screen has been calculated. Fig. 1 shows the geometry considered and the symbols used. The field strength of DR for the slit $\vec{E}_{slit}$ is considered as a superposition of DR fields from the upper ($\vec{E}_u$) and lower ($\vec{E}_d$) semi-planes:

$$\vec{E}_{slit} = \vec{E}_u e^{i\varphi_u} + \vec{E}_d e^{-i\varphi_d} \qquad (1)$$

Here $\varphi_u$, $\varphi_d$ are the phase shifts calculated for the upper and lower slit edges relative to the particle trajectory. The total phase shift $\varphi_0$ can be derived from simple geometrical relations (see Fig.1):

$$\varphi_0 = \varphi_u + \varphi_d = \frac{2\pi a}{\lambda}(\cos(\theta - \theta_0) - \frac{\cos\theta_0}{\beta}) =$$
$$= \frac{2\pi(h_1 + h_2)}{\lambda \sin\theta_0}\left(\cos(\theta - \theta_0) - \frac{\cos\theta_0}{\beta}\right) \qquad (2)$$

In (2) λ is the DR wavelength and β is the particle velocity. Here and further the system of units $\hbar = m = c = 1$ is used. In the paper [2] it is shown that DR from a relativistic particle is concentrated in the cone $\sim \gamma^{-1}$ near the specular direction. In the coordinate system, where the Z axis coincides with that direction and the X axis is in parallel to the slit edge, from (2) one can derive (up to an accuracy of the terms of $\gamma^{-2}$):

$$\varphi_u \approx -\frac{2\pi h_1}{\lambda}\theta_y \ , \quad \varphi_d \approx -\frac{2\pi h_2}{\lambda}\theta_y \tag{3}$$

After substituting in (1) $\vec{E}_u$ and $\vec{E}_d$ values obtained in the same approximation [2], we have BDR field for the slit:

$$E_{x,slit} = \frac{iZe}{4\pi^2}\frac{\theta_x}{\sqrt{\gamma^{-2}+\theta_x^2}}\left(\frac{\exp\left[-\frac{2\pi h_1}{\lambda}\left(\sqrt{\gamma^{-2}+\theta_x^2}+i\theta_y\right)\right]}{\sqrt{\gamma^{-2}+\theta_x^2}+i\theta_y} + \frac{\exp\left[-\frac{2\pi h_2}{\lambda}\left(\sqrt{\gamma^{-2}+\theta_x^2}-i\theta_y\right)\right]}{\sqrt{\gamma^{-2}+\theta_x^2}-i\theta_y}\right),$$

$$E_{y,slit} = \frac{Ze}{4\pi^2}\left(-\frac{\exp\left[-\frac{2\pi h_1}{\lambda}\left(\sqrt{\gamma^{-2}+\theta_x^2}+i\theta_y\right)\right]}{\sqrt{\gamma^{-2}+\theta_x^2}+i\theta_y} + \frac{\exp\left[-\frac{2\pi h_2}{\lambda}\left(\sqrt{\gamma^{-2}+\theta_x^2}-i\theta_y\right)\right]}{\sqrt{\gamma^{-2}+\theta_x^2}-i\theta_y}\right) \tag{4}$$

In (4) Ze is the particle charge. Expressions (4) coincide with the results of M. L. Ter-Mikaelian [7], obtained for FDR at Z=1. The coincidence of BDR and FDR characteristics is connected with the use of the ideally conducting screen approximation. BDR spectral – angular distribution is simply calculated from (4)

$$\frac{d^2W}{d\omega d\Omega} = 4\pi^2\left(|E_x|^2+|E_y|^2\right) =$$
$$= \frac{Z^2\alpha}{4\pi^2}\gamma^2 \frac{\exp\left(-\frac{\omega}{\omega_c}\sqrt{1+t_x^2}\right)}{(1+t_x^2)(1+t_x^2+t_y^2)}\left\{(1+2t_x^2)\left[\exp\left(2l\frac{\omega}{\omega_c}\sqrt{1+t_x^2}\right)+\exp\left(-2l\frac{\omega}{\omega_c}\sqrt{1+t_x^2}\right)\right] - \right. \tag{5}$$
$$\left. -\frac{2}{(1+t_x^2+t_y^2)}\left[(1+t_x^2-t_y^2)\cos\left(\frac{\omega}{\omega_c}t_y\right)-2t_y\sqrt{1+t_x^2}\sin\left(\frac{\omega}{\omega_c}t_y\right)\right]\right\}$$

Here α is the fine structure constant, $t_x = \gamma\theta_x$, $t_y = \gamma\theta_y$, $\varpi_c = \gamma/(a\sin\theta_0) = \gamma/(h_1+h_2)$ is the characteristic energy of DR, $l = \frac{h_1-\frac{a}{2}\sin\theta_0}{a\sin\theta_0}$ is the relative impact parameter of the particle with respect to the slit center. As in the case of OTR, in the BDR distribution there is a zero minimum at $t_x = t_y = 0$ when the particle moves through the slit's center. It is obvious that the optical BDR angular distribution shape is connected with the initial particle beam divergence (the same as in the case of OTR).

If the target with a slit is placed on a goniometer and the BDR emitted at the angle $\theta$ is detected by means of collimator C, optical filter F and photomultiplier D (see Fig.1), then with changing of the target tilted angle $\theta_0$ one can measure the dependence of BDR yield (theta-scan), which is rather close to BDR angular distribution. Indeed, when target rotates on a value $\Delta\theta_0$, specular direction (Z axis) removes on the value $2\Delta\theta_0$. Thus, putting $t_y \to 2\gamma\Delta\theta_0$ in (5) and integrating over the angular variable $t_x$ in the range of the collimator aperture $\Delta\theta_x$, we obtain the theta-scan distribution. The energy resolution of the equipment, distribution of the initial beam impact parameters, the beam angular divergence etc. can be taken into account by the fully obvious way. The theta-scan distribution simulation for the beam, which divergence is described by the Gaussian:

$$F(\Delta_x, \Delta_y) = \frac{1}{2\pi\sigma^2} \exp\left(-\frac{\Delta_x^2 + \Delta_y^2}{2\sigma^2}\right) \qquad (6),$$

has been carried out by the following way. In (6) $\Delta_x$, $\Delta_y$ are the deviation angles of the initial particle from the average direction. For simplification of the calculations the collimator was chosen for $\Delta\theta_y \leq \gamma^{-1}$, $\Delta\theta_x \gg \gamma^{-1}$. The convolution of the distributions (5) and (6) with integration over $\theta_x$ in the infinite region allows us to obtain a one-dimensional theta-scan distribution for $\varpi = 0.5\varpi_c$ (see Fig.2). As it follows from the figure, for the beam divergence $\sigma \geq \gamma^{-1}$ the resulting distribution has a single maximum, instead of the abovementioned minimum at $t_y = 0$. Thus, measuring a similar theta-scan one can obtain the parameter $\sigma$ describing an isotropic initial beam divergence. For a non-isotropic distribution (i.e. $\sigma_x \neq \sigma_y$) it is necessary either to measure the theta-scan for a narrow aperture ($\Delta\theta_x$, $\Delta\theta_y \sim \gamma^{-1}$) or to measure it during rotation of the target with the slit around a perpendicular axis. When an electron beam the with $\gamma = 2000$ moves through a slit with the width a = 0.3mm inclined at the angle $\theta_0 = 45°$, the BDR photon yield in the optical region is only $e$ times less than the OTR photon yield for the same conditions. However, in the latter case beam emittance increases due to multiple scattering in the target, which is absent in the case of DR.

Let us consider the DR characteristics for non-relativistic particles ($\gamma \sim 1$, $\beta \ll 1$). In that case the expression for spectral – angular density can be obtained from the exact result [1]. For non-relativistic particle angular distribution is quite broad and is not determined by particle velocity. Omitting cumbersome derivation, we present the formula for the DR spectrum (neglecting the terms of the order $\sim \beta^2$):

$$\frac{dW}{d\omega} = \frac{3}{4} Z^2 \alpha\beta \exp\left(-\frac{2\omega h}{\beta}\right) \qquad (7)$$

Here h is the impact parameter. In comparison with the DR from relativistic electrons accelerated nuclear beam emits $Z^2$ times more intensively, however, the factor $\beta \cdot \exp(-2\varpi h/\beta)$ leads to significant suppression of the DR yield in the optical region even for micron-size beams (in that case the impact parameter can also be chosen of a micron size). However, in the infrared region ($\lambda \sim 10\, mcm$) the radiation yield for the beam of protons with $\gamma = 2$ ($\beta = 0.87$) can be measured if the beam diameter and the impact parameter do not exceed 1mcm. In that case the photon yield in an interval of $\Delta\lambda/\lambda = 0.1$ and in the solid angle $\Delta\Omega = 0.1\, sterad$ can reach $\sim 10^{-6}$ photon/particle, that allows one to discuss the feasibility of using BDR for non-destructive diagnostics of both relativistic electron beams and beams of protons and nuclei.

**Acknowledgements**
The authors are thankful to Prof. I.P.Chernov for useful discussions and P.V.Karataev for help in preparing the text of the paper.

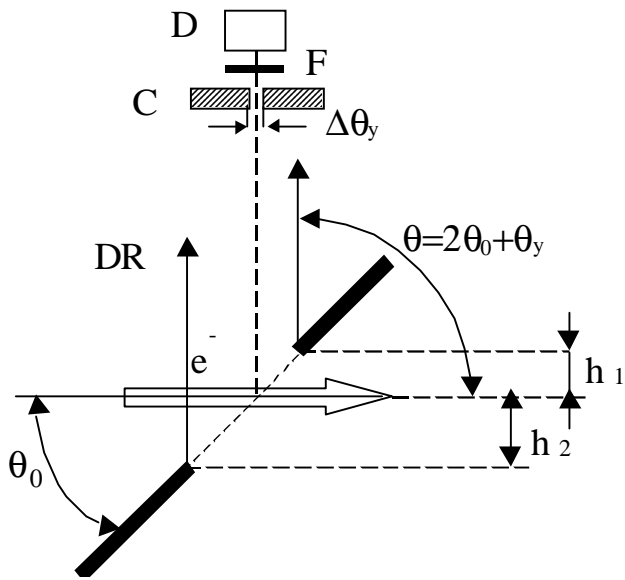

**Fig.1**  Layout of the measurements of BDR yield with changing of the tilted angle $\theta_0$ (theta-scan).

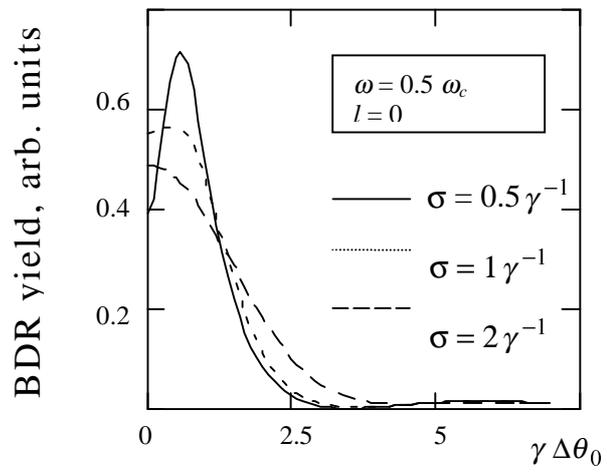

**Fig.2** Theta-scan distributions for detector placed at the fixed angle $\theta = 2\theta_0$ for beams with different divergence $\sigma$.